# Oxygen non-stoichiometry in Ru-1212 and Ru-1222 magnetosuperconductors

M. Matvejeff[a,b], V.P.S. Awana[a], L.-Y. Jang[c], R.S. Liu[d], H. Yamauchi[a] and M. Karppinen[a,*]

[a]*Materials and Structures Laboratory, Tokyo Institute of Technology, Yokohama 226-8503, Japan*

[b]*Laboratory of Inorganic and Analytical Chemistry, Helsinki University of Technology, FIN-02015 Espoo, Finland*

[c]*Synchrotron Radiation Research Center, Hsinchu, Taiwan, R.O.C.*

[d]*Department of Chemistry, National Taiwan University, Taipei, Taiwan, R.O.C.*

**Abstract**

Here we report the results of thermogravimetric (TG) analysis on the oxygen non-stoichiometry of $RuSr_2GdCu_2O_{8-\delta}$ (Ru-1212) and $RuSr_2(Gd_{0.75}Ce_{0.25})_2Cu_2O_{10-\delta}$ (Ru-1222) samples. With TG annealings carried out in $O_2$ and Ar atmospheres it was found that the oxygen content in Ru-1212 remains less affected upon various annealings, while for Ru-1222 wider-range oxygen-content tuning is possible. When heated in $H_2$/Ar atmosphere the both phases release oxygen upon breaking down to mixtures of metals (Ru and Cu) and binary oxides ($CeO_2$, $Gd_2O_3$, and SrO) in two distinct steps around 300 and 450 °C. This reductive decomposition reaction carried out in a thermobalance was utilized in precise oxygen content determination for these phases. It was found that the 100-atm $O_2$-annealed Ru-1212 sample was nearly stoichiometric, while the similarly treated Ru-1222 sample was clearly oxygen deficient. X-ray absorption near-edge (XANES) spectroscopy was applied to estimate the valence of Ru in the samples. In spite of the fact that the Ru-1212 phase was shown to possess less oxygen-deficient $RuO_{2-\delta}$ layer, the valence of Ru as probed with XANES was found to be lower in Ru-1212 than that in Ru-1222.

*Corresponding author

Prof. M. Karppinen

Materials and Structures Laboratory, Tokyo Institute of Technology, 4259 Nagatsuta, Midori-ku, Yokohama 226-8503, Japan

Phone: +81-45-924-5333

Fax: +81-45-924-5365

E-mail address: karppinen@msl.titech.ac.jp

## I. Introduction

Since the first reports on the coexistence of high-$T_c$ superconductivity and ferromagnetism in samples of $RuSr_2GdCu_2O_{8-\delta}$ (Ru-1212) [1] and $RuSr_2(Gd_{0.75}Ce_{0.25})_2Cu_2O_{10-\delta}$ (Ru-1222) [2], these "magnetosuperconductor" phases have received continuous attention. However, despite the intensive research made on the both phases, their basic characteristics in terms of oxygen non-stoichiometry, carrier concentration and valence state of Ru are not understood well yet. In the present contribution we present the results of a systematic study on the oxygen non-stoichiometry in Ru-1212 and Ru-1222 samples based on thermogravimetric (TG) analyses carried out in $O_2$, Ar and $H_2$/Ar atmospheres. The experiments in $O_2$ and Ar atmospheres reveal us the degree of oxygen content tunability in these phases. The experiments in $H_2$/Ar, on the other hand, enable us to determine the absolute oxygen content of the sample. Furthermore, we apply Ru $L_{III}$-edge x-ray absorption near-edge structure (XANES) spectroscopy to address the question on the valence state of Ru in the Ru-1212 and Ru-1222 phases. Prior to the present study, the very same samples have undergone thorough characterizations for structural, magnetic and superconductivity properties. The results of these studies have been reported elsewhere [3-6].

The structure of Ru-1212 is derived from that of $R$Ba$_2$Cu$_3$O$_{7-\delta}$ or CuBa$_2R$Cu$_2$O$_{7-\delta}$ (Cu-1212; $R$ = rare earth element) with Cu in the charge reservoir replaced by Ru, such that the CuO$_{1-\delta}$ chain is replaced by a RuO$_{2-\delta}$ sheet. The Ru-1222 structure, on the other hand, results from that of Ru-1212 when inserting a ($R$,Ce)-O$_2$-($R$,Ce) fluorite-type block instead of the single oxygen-free $R$ layer, between the two adjacent CuO$_2$ planes. The Ru-1222 phase is thus recognized as an $M$ = Ru, $m$ = 1, $k$ = 1, $s$ = 2 member of "category-B" cuprates with a general layer sequence of $A$O-($M$O$_{1\pm\delta/m}$)$_m$-$A$O-CuO$_2$-$B$-[O$_2$-$B$]$_{s-1}$-CuO$_2$ and a chemical formula of $M_mA_{2k}B_s$Cu$_{1+k}$O$_{m+4k+2s\pm\delta}$ or $M$-$m(2k)s(1+k)$ [7]. Now the question is whether the oxygen stoichiometry of the Ru-1212 and Ru-1222 phases can be controlled in the same continuous manner as that well established for Cu-1212 [7,8].

**II. Experimental**

Samples of RuSr$_2$GdCu$_2$O$_{8-\delta}$ (Ru-1212) and RuSr$_2$(Gd$_{0.75}$Ce$_{0.25}$)$_2$Cu$_2$O$_{10-\delta}$ (Ru-1222) were synthesized through a solid-state reaction route from stoichiometric amounts of RuO$_2$, SrO$_2$, Gd$_2$O$_3$, CeO$_2$ and CuO. Calcinations were carried out at 1000 °C, 1020 °C and 1040 °C for 24 hours at each temperature with intermediate grindings. The powder samples thus obtained were pelletized and annealed in a flow of oxygen at 1075 °C for 40 hours with a slow cooling over a span of 20 hours down to room temperature. A portion of these as-synthesized samples was further annealed in 100 atm O$_2$ atmosphere for 100 hours at 420 °C and subsequently cooled slowly to room temperature in the same atmosphere. The phase purity and lattice parameters of the synthesized samples were checked by x-ray diffraction (XRD) measurements (MAC Science: MXP18VAHF[22]; Cu$K_\alpha$ radiation). The $T_c$ values reported for the superconductivity transition seen in the samples are from magnetization measurements carried out with a SQUID magnetometer (Quantum Design: MPMS-XL) [3-6].

The TG analyses (Perkin Elmer: TAS-7) were carried out in Ar, $O_2$ and 5% $H_2$/95% Ar atmospheres. The mass of the sample was ca. 15 mg and heating rates of both 1 and 5 °C/min were applied. For the experiments carried out in $O_2$ atmosphere not only the heating curves but also the cooling curves were recorded to investigate the reversibility of the oxygen depletion/uptake process. The Ru $L_{III}$-edge XANES measurements were performed at the BL15B beamline of the Synchrotron Radiation Research Center (SRRC) in Hsinchu, Taiwan. Details of the experimental procedure used were described earlier [9].

### III. Results and Discussion

As commonly seen in the reported XRD spectra of Ru-1212 samples, also for the present Ru-1212 samples a small amount of $SrRuO_3$ was detected as an impurity phase. The Ru-1222 phase was more easily obtained in single-phase form, although a very trace of $SrRuO_3$ was distinguished from the XRD patterns even for the Ru-1222 samples. For the both phases, the relative amount of $SrRuO_3$ in the samples was considered to be small enough (< 5 mass-%) not to affect the results of TG and XANES analyses.

The lattice parameters for the as-synthesized Ru-1212 sample were determined in space group $P4/mmm$ at: $a = b = 3.821(1)$ Å and $c = 11.476(1)$ Å. Essentially no difference was found in lattice parameters for the 100-atm $O_2$-annealed Ru-1212 sample, suggesting that the oxygen content did not change considerably upon the high-$O_2$-pressure annealing. Also the $T_c$ remained unchanged at 20 K. For the Ru-1222 samples the lattice parameters were determined in space group $I4/mmm$. The obtained values are: $a = b = 3.834(1)$ Å and $c = 28.493(1)$ Å for the as-synthesized sample and $a = b = 3.833(1)$ Å and $c = 28.393(1)$ for the 100-atm $O_2$-annealed sample. The slightly decreased $c$-axis parameter of the 100-atm $O_2$-annealed Ru-1222 sample as compared to that of the as-synthesized sample indicates the possibility of successful introduction of extra oxygen into the Ru-1222 structure upon the high-pressure $O_2$-annealing. Accordingly the value of $T_c$ increased from 23 K to 43 K.

When heating the Ru-1222 samples in a thermobalance, they exhibited definite changes in weight, apparently caused by changes in the oxygen content. The oxygen non-stoichiometry of the Ru-1222 samples was therefore systematically studied with various heat treatments in $O_2$ and Ar atmospheres. A representative TG curve obtained for the 100-atm $O_2$-annealed Ru-1222 sample upon Ar annealing is shown in Fig. 1(a). The total amount of oxygen *per* formula unit ($\Delta\delta$) found to be depleted from the structure, before the phase decomposed above 900 °C was calculated at $\Delta\delta = 0.28(3)$. Here it should be recognized that the TG curve recorded in Ar atmosphere for an (100-atm) $O_2$-annealed sample allows one to read the proper annealing temperatures (corresponding to the desired oxygen contents) when aiming at preparing (a series of) samples with fixed, intermediate oxygen-content values, *i.e.* so-called TCOD (temperature-controlled oxygen depletion) annealing [7,8].

Upon heating in $O_2$ the Ru-1222 samples release oxygen as well. The amount of oxygen released from the 100-atm $O_2$-annealed sample upon heating in $O_2$ by 900 °C ($\Delta\delta = 0.27(3)$) was almost the same as for the treatment in Ar. During the slow (1 °C/min) cooling in $O_2$ from 900 °C back to room temperature the sample re-accommodated part of the oxygen released from it during the heating. However, not all of the oxygen lost in heating was taken back to the sample during the cooling. This is due to the fact that the initial annealing of the sample had been performed in 100-atm $O_2$, while the TG measurement was performed in 1 atm $O_2$. The difference in oxygen content between the 100-atm and 1-atm $O_2$-annealed samples was estimated at 0.16(3) from the obtained TG curve for the heating and cooling cycles.

Experiments performed for the Ru-1212-phase samples in a thermobalance in $O_2$ and Ar indicated that, as compared to the Ru-1222 phase, the oxygen in the Ru-1212 structure is relatively immovable and only small changes in the oxygen stoichiometry were seen with heat treatments carried out in $O_2$ and Ar atmospheres. In Fig. 1(b), a representative TG curve for an Ar annealing of as-synthesized Ru-1212 sample is presented.

The absolute oxygen contents of the samples of both Ru-1212 and Ru-1222 phases were determined based on a TG heat treatment in a strongly reductive 5% $H_2$/95% Ar atmosphere. We present examples of such TG curves for the 100-atm $O_2$-annealed Ru-1222 sample and the as-synthesized Ru-1212 sample in Figs. 2(a) and 2(b), respectively. For both the phases the decomposition of sample occurs in two distinct steps about 200-350 $^o$C and 400-500 $^o$C. To clarify the process of the sample decomposition during the different steps of reduction, additional TG measurements were performed in 5% $H_2$/95% Ar atmosphere for the simple oxides of the constituent metals, *i.e.* $RuO_2$, $CuO$, $Cu_2O$ and $CeO_2$. The TG curves obtained are shown in Fig. 3. It is seen that $RuO_2$, $CuO$ and $Cu_2O$ decompose at low temperatures in a single step reaction at about 100 $^o$C, 150 $^o$C and 330 $^o$C, respectively, while $CeO_2$ remains stable at least up to 750 $^o$C. We therefore calculated the exact oxygen contents of the Ru-1212 and Ru-1222 samples from the weight losses seen in the 5% $H_2$/95% Ar reduction curves by 550 $^o$C assuming the final decomposition product to be a mixture of binary oxides, $SrO$, $Gd_2O_3$ and $CeO_2$, and metals, Ru and Cu. The results are presented in Table 1, implying that the 100-atm $O_2$-annealed Ru-1212 sample is stoichiometric within the error bars of the analysis, while the 100-atm $O_2$-annealed Ru-1222 sample is clearly oxygen deficient. Furthermore, as already predicted based on the XRD data (lattice parameters) and the $T_c$ values, the difference between the 100-atm and 1-atm $O_2$-annealed samples is larger for the Ru-1222 phase than for the Ru-1212 phase, see also the data for relevant samples in Refs. 12 and 13. In Table 1, we also show the absolute oxygen content values for the Ru-1212 and Ru-1222 phases annealed in Ar up to the highest temperatures before the break-down of the structures, *i.e.* 750 $^o$C for Ru-1212 and 900 $^o$C for Ru-1222. These values, *i.e.* 7.80(5) for Ru-1212 and 9.35(5) for Ru-1222, represent the minimum oxygen contents tolerated by the phases.

The Ru $L_{III}$-edge XANES spectra were recorded for the as-synthesized Ru-1212 sample and the as-synthesized and 100-atm $O_2$-annealed Ru-1222 samples. The obtained spectra were analyzed quantitatively by fitting them to certain linear combinations of spectra for reference materials for $Ru^{IV}$ ($Sr_2RuO_4$) and $Ru^V$ ($Sr_2GdRuO_6$). This fitting approach was first applied in Refs. 10 and 11. For the two Ru-1222 samples the spectra and fitting were shown in Ref. 9. Here we show only the spectrum of the as-synthesized Ru-1212 sample in Fig. 4. All the three samples are between the two reference materials in terms of the ruthenium valence. Fitting the spectra revealed a valence value of +4.74(5) for the as-synthesized Ru-1222 sample and +4.81(5) for the 100-atm $O_2$-annealed Ru-1222 sample [9]. It seems that the valence of Ru in Ru-1222 depends on the oxygen content, thus indirectly suggesting that the changes in oxygen stoichiometry occur in the $RuO_{2-\delta}$ layer. For the as-synthesized Ru-1212 sample a valence value of +4.62(5) was obtained. This value is very close to that reported in Ref. [10] for another $RuSr_2GdCu_2O_{8-\delta}$ sample, *i.e.* +4.60, with the final annealing performed in $O_2$ at 1060 °C.

Though it would be tempting to do so, we do not calculate any precise valence values for Cu from the obtained oxygen content and Ru valence values. This is due to the fact that considering the error bars for our TG and XANES analyses, we believe that such estimation would end up to not-reliable-enough Cu valence values. However, rough assessment suggests that the valence of Cu is lower in the Ru-1222 phase samples than in the samples of the Ru-1212 phase.

Finally, the considerably larger oxygen non-stoichiometry seen for the Ru-1222 phase, as compared to the Ru-1212 phase, might let us to believe that not only the $RuO_{2-\delta}$ sheet but also the $(Gd_{0.75}Ce_{0.25})$-$O_2$-$(Gd_{0.75}Ce_{0.25})$ fluorite block in Ru-1222 is non-stoichiometric in terms of oxygen content. However, for the related Co-1222 phase, we recently confirmed that the ($Y_{0.75}Ce_{0.25}$)-$O_2$-($Y_{0.75}Ce_{0.25}$) fluorite block in it is oxygen-stoichiometric even when exposed to various reductive and oxygenative annealings [14].

## IV. Conclusion

In the present study we have shown that the Ru-1212 phase is easily obtained stoichiometric, while the Ru-1222 phase was clearly oxygen-deficient even after 100-atm $O_2$-annealing. Also was shown that the oxygen content in the Ru-1212 phase remains less affected upon various annealings, while for the Ru-1222 sample wider-range oxygen-content tuning is possible. We also presented a method for fine-tuning the oxygen non-stoichiometry in those phases, which can be used to prepare sample series with gradually changing oxygen content.

**Acknowledgement**

The present work has been supported by a Grant-in-Aid for Scientific Research (Grant No. 11305002) from the Ministry of Education, Science and Culture of Japan and by the Academy of Finland (Decision No. 46039). M. Matvejeff acknowledges Finnish Cultural Foundation, Sasakawa Foundation and the Komppa Foundation for grants for PhD studies.

**Table 1.** The value of oxygen content as determined from the TG analysis for the 100-atm $O_2$-annealed, 1-atm $O_2$-synthesized and 1-atm Ar-annealed Ru-1212 and Ru-1222 samples. The values for the 100-atm $O_2$-annealed and 1-atm $O_2$-synthesized samples are from the $H_2$/Ar reductions, while those for the 1-atm Ar-annealed samples are calculated from the TG experiments carried out in Ar atmosphere at temperatures close to the decomposition temperatures of the two phases. These latter values represent the minimum oxygen contents tolerated by the two phases. The value of superconductivity transition temperature, $T_c$, is given in parentheses when being measured.

| Synthesis/annealing conditions | Oxygen content (*per* formula unit) | |
| --- | --- | --- |
| | Ru-1212 | Ru-1222 |
| 100-atm $O_2$, 420 °C | 7.98(5) ($T_c$ = 20 K) | 9.63(5) ($T_c$ = 43 K) |
| 1-atm $O_2$, 1075 °C (as-synthesized) | 7.93(5) ($T_c$ = 20 K) | 9.54(5) ($T_c$ = 23 K) |
| 1-atm Ar, 750/900 °C (Ru-1212/Ru-1222) | 7.80(5) | 9.35(5) |

**FIGURE CAPTIONS**

**Fig. 1.** TG curves recorded for a) the 100-atm $O_2$-annealed Ru-1222 sample, and a) the as-synthesized Ru-1212 sample in Ar atmosphere. The heating rate was 1 °C/min and the mass of the sample *ca*. 15 mg.

**Fig. 2.** TG curves recorded for a) the 100-atm $O_2$-annealed Ru-1222 sample, and b) the as-synthesized Ru-1212 sample in 95 % Ar/5 % $H_2$ atmosphere. The heating rate was 1 °C/min and the mass of the sample *ca*. 15 mg.

**Fig. 3.** TG curves recorded for a) $RuO_2$, b) CuO, c) $Cu_2O$, and d) $CeO_2$ in 95 % Ar / 5 % $H_2$ atmosphere. The heating rate was 1 °C/min and the mass of the sample *ca*. 15 mg.

**Fig. 4.** Ru $L_{III}$-edge XANES spectrum collected for the as-synthesized Ru-1212 sample. Also shown are the spectra for reference samples $Sr_2RuO_4$ ($Ru^{IV}$) and $Sr_2GdRuO_6$ ($Ru^V$).

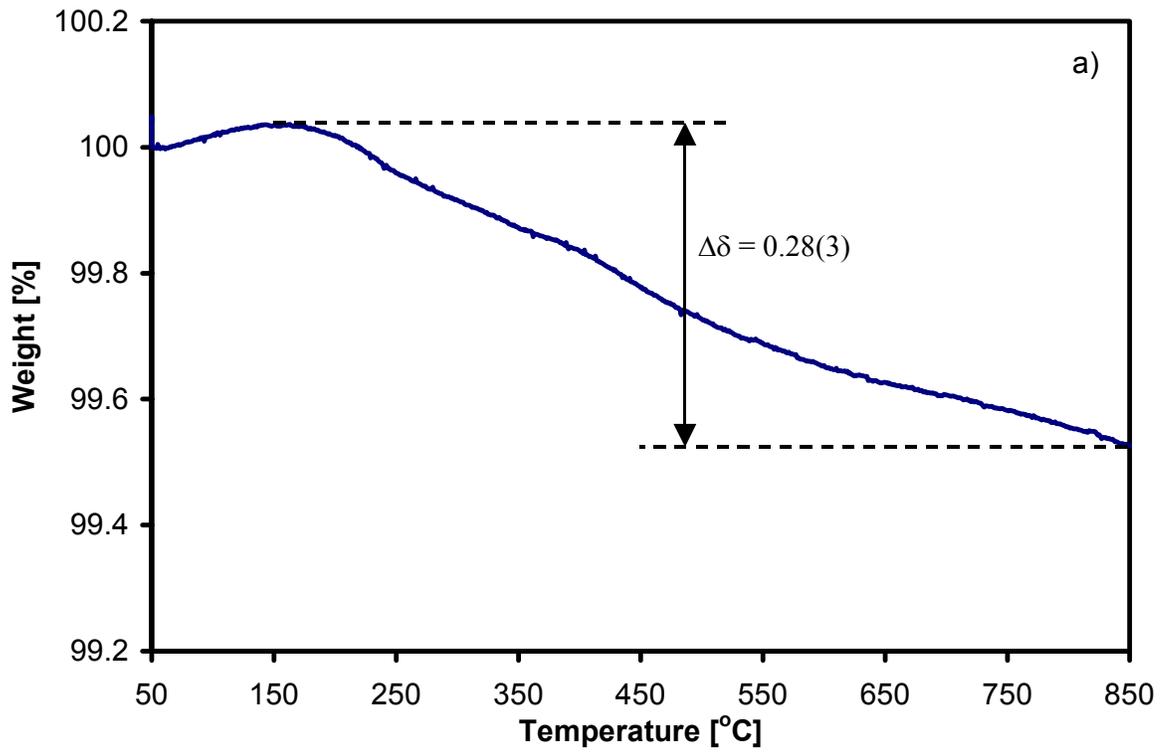

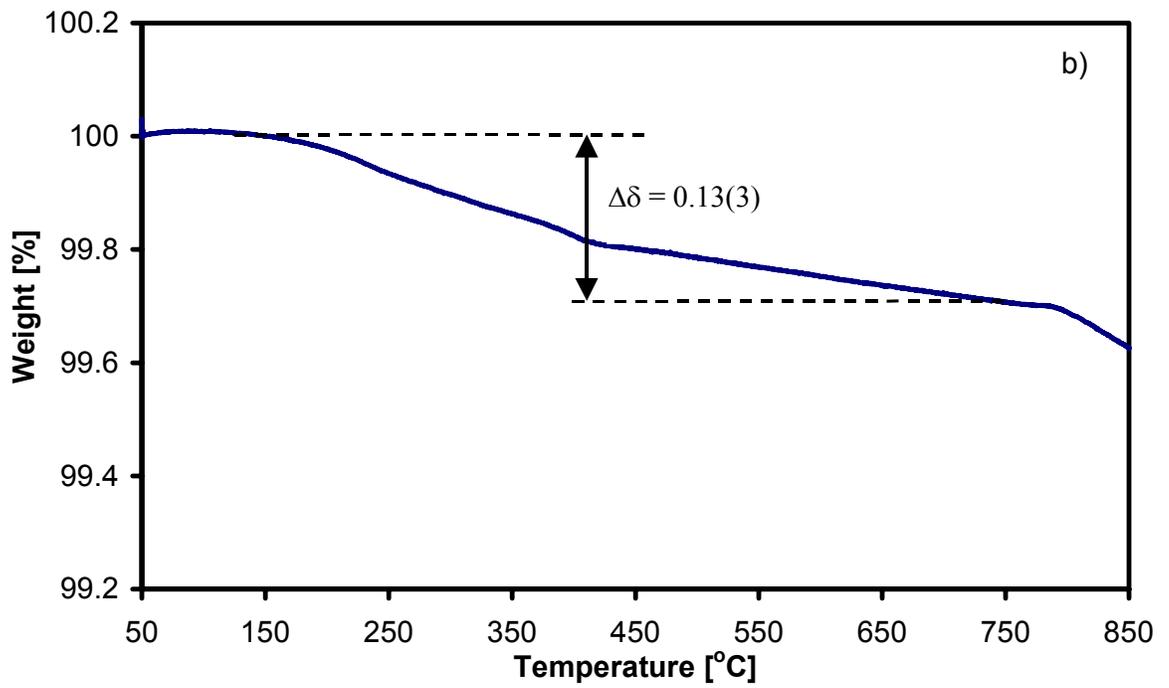

**Matvejeff** *et al.*: Fig. 1.

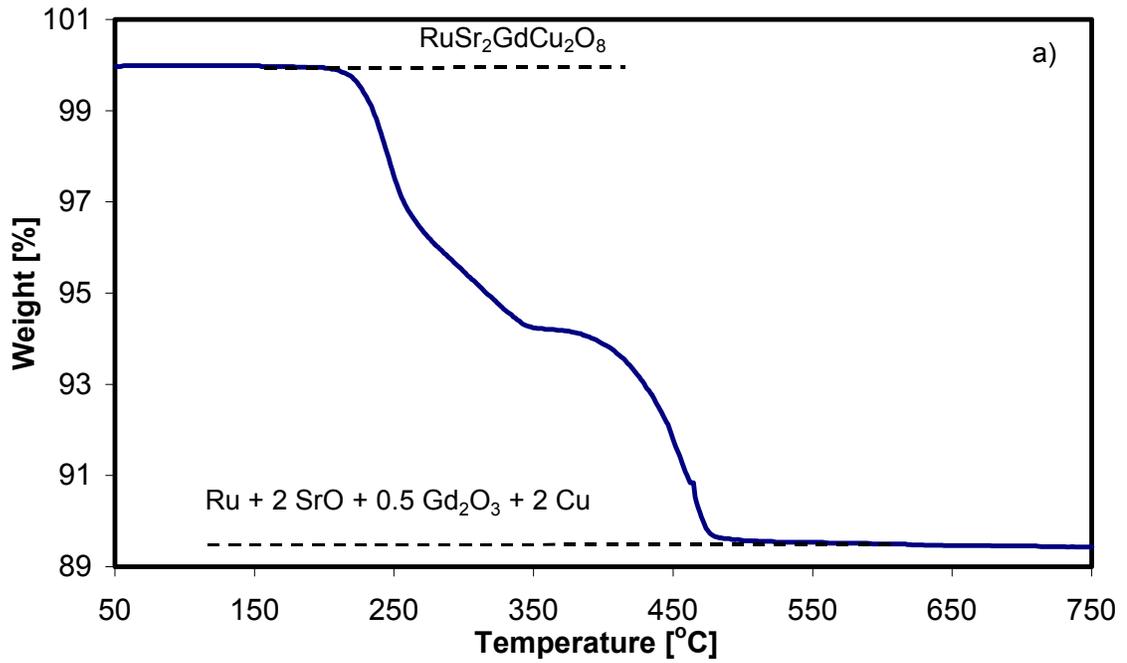

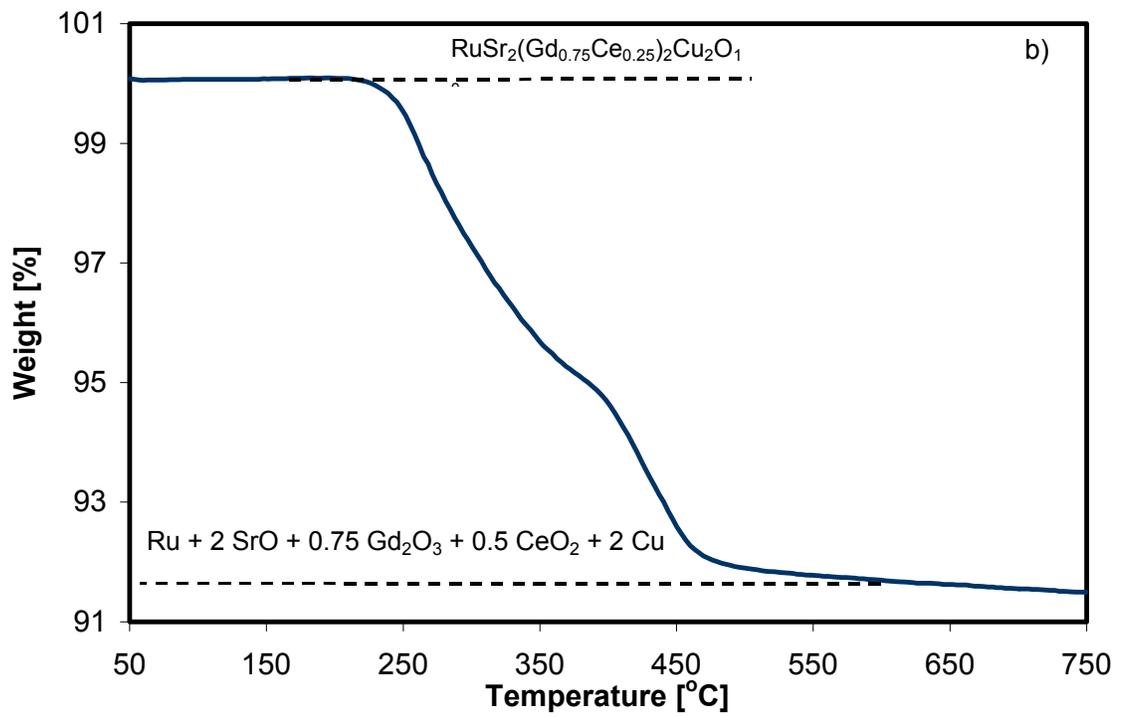

**Matvejeff *et al.*: Fig. 2.**

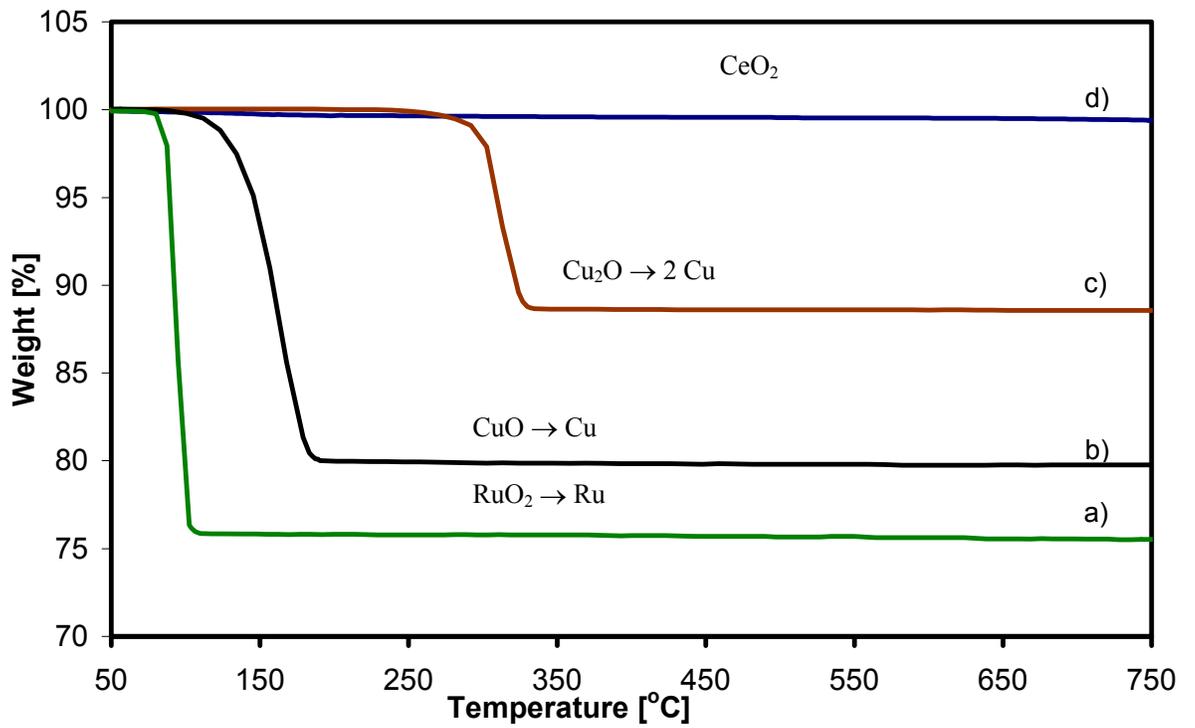

**Matvejeff** *et. al.*: **Fig. 3**

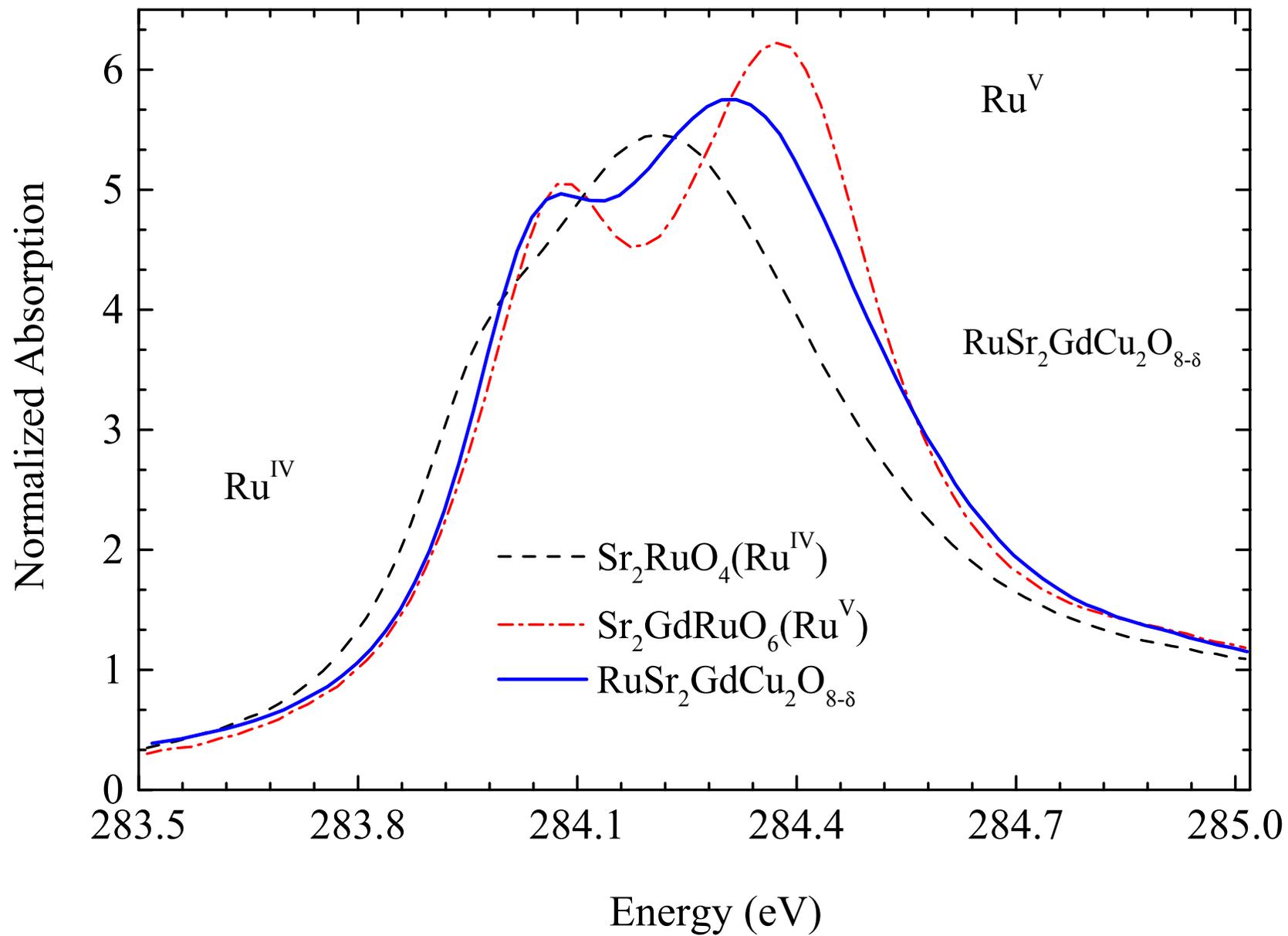